# Tidal Frequencies in the Time Series Measurements of Atmospheric Muon Flux from Cosmic Rays

H. Takai[1,*], C. Feldman[2], M. Minelli[3], J. Sundermier[4], G. Winters[5], M. K. Russ[5,6], J. Dodaro[5,7], A. Varshney[5,8], C. J. McIlwaine[5,9], T. Tomaszewski[10], J. Tomaszewski[11], R. Warasila[12], J. McDermott[13], U. Khan[13], K. Chaves[7], O. Kassim[14], and J. Ripka[15]

[1]*Physics Department, Brookhaven National Laboratory, Upton NY 11973*
[2]*Department of Physics and Astronomy, Stony Brook University, Stony Brook, NY 11974*
[3]*Physics Department, College of the Holy Cross, Worcester, MA 01610*
[4]*Deer Park High School, Deer Park, NY 11729*
[5]*Smithtown High School East, Smithtown, NY 11780*
[6]*Toshiba Stroke and Vascular Res. Center, University at Buffalo (SUNY), Buffalo, NY 14260*
[7]*Physics Department, Stanford University, Stanford, CA 94305*
[8]*NYIT College of Osteopathic Medicine, Old Westbury, NY 11568*
[9]*Galileo Global Advisors, LLC, New York, NY 10020*
[10]*Shoreham-Wading River High School, Shoreham, NY 11786*
[11]*Science Department, Haverford High School, Haverford, PA 19083*
[12]*Suffolk County Community College, Ammerman Campus, Selden, NY 11784*
[13]*William Floyd High School, Mastic Beach, NY 11951*
[14]*School of Architecture, Syracuse University, Syracuse, NY 13244*
[15]*MacArthur High School, Levittown, NY 11756*

[*]Correspondence to: takai@bnl.gov

**Abstract**: Tidal frequencies are detected in time series muon flux measurements performed over a period of eight years. Meson production and subsequent decay produce the muons that are observed at ground level. We interpret the periodic behavior as a consequence of high altitude density variations at the point of meson production. These variations are driven by solar thermal cycles. The detected frequencies are in good agreement with published tidal frequencies and suggest that muons can be a complementary probe to the study of atmospheric tides at altitudes between 20 to 60 *km*.

**One Sentence Summary:** Tidal frequencies are detected in a time series measurement of muon flux over a period of eight years.

Cosmic rays are high energy atomic nuclei that are produced and accelerated by astrophysical events such as supernovae *(1,2)*. Upon entering Earth's atmosphere, they interact with air molecules, producing a shower of elementary particles whose multiplicity and species population depends on the energy of the incident cosmic ray. The first interaction takes place at altitudes ~50-60 km above sea level, with subsequent interactions happening at lower altitudes, most frequently at ~20 km. Within the particle shower, muons ($\mu^\pm$) are the decay products of mesons, e.g. pions ($\pi^0$, $\pi^\pm$) and kaons ($K^0$, $K^\pm$). To be detected at ground level, a muon must live long enough to travel the flight path from its point of creation to the detector. Changes in atmospheric height such as those caused by the seasonal expansion and contraction of Earth's atmosphere move the muon creation point higher or lower. This variation is observed as a seasonal dependence in the detected muon rate *(3,4,5)*. At lower altitudes, where particle production ceases, the denser atmosphere can bring muons to a stop, preventing them from reaching the ground level. This is best shown by the anti-correlation between the number of muons detected and the barometric pressure, used as a proxy for the atmospheric density *(6)*.

There are several mechanisms that lead to changes in atmospheric density. Solar heating of atmospheric layers is the dominant mechanism, and gives origin to atmospheric tides *(7-10)*. These tides are oscillatory movements of air masses characterized by a set of accurately known frequencies that reflect the amount of daily insolation as Earth revolves around the Sun. The Earth's tilted rotation axis with respect to the ecliptic compounded with its variable orbital velocity makes the amount of insolation a phase modulated time function *(11)*.

We report on the detection of tidal frequencies in the spectral analysis of time series muon flux measurements realized over a period of eight years. The muon telescope used for these measurements is part of the MARIACHI experiment, and is located at Smithtown High School East in the state of New York, latitude 40° 52' 14.88"N, longitude 73° 9' 53.103"W, at 43 *m* above sea level *(12)*. The detection system consists of two 0.28 $m^2$ plastic scintillators subtending a solid angle of 3.8 *sr*. Counts per minute are recorded by a computer and assigned a time stamp provided by a global positioning system (GPS) clock that has a nominal accuracy of 100 *ns*. The setup is located indoors with ~19 $g/cm^2$ of roofing material above the detectors and detects muons with momenta above ~200 *MeV/c*. Data is unevenly spaced due to power outages, equipment failure, and unavailability of the data acquisition computers, resulting in gaps in data collection. The total number of samples recorded is $3.391 \times 10^6$, about 80% of the expected $4.205 \times 10^6$ if data collection was not disrupted. The average measured muon rate is $(1890 \pm 51)$ *counts/min*, which translates into a rate of $(29.6 \pm 0.8)$ *counts/s $m^2$ sr*. The standard deviation indicates the flux variation around the average.

Prior to performing the frequency analysis, we averaged values over each hour and then performed barometric pressure corrections using readings from a weather station located at the MacArthur airport, 12 *km* south of the high school. This correction compensates the measured flux for losses due to density variations in the lower layers of the troposphere. The barometric coefficient found is $(1/P)(\Delta n/n) = 0.17\%$ $hPa^{-1}$, which is close to those obtained from experiments performed with low energy thresholds and at low altitudes *(13,14)*.

The hourly-averaged, pressure corrected time series data used for this analysis are depicted in Fig. 1 (top). The most striking feature is the yearly modulation with an amplitude of ±5% of the average counts, with maxima and minima during winter and summer seasons, respectively. This modulation is caused by seasonal variations in solar heating that expand or contract the atmosphere. As previously mentioned, this change in atmospheric thickness lengthens or contracts the muon flight path. An inspection of the data also shows several dips in counts that correspond to Forbush deficit events caused by solar coronal mass ejections *(15)*. The largest of these dips was observed in February of 2014. The data also show periods of increase in the muon counts of unknown origin.

The spectral analysis of our data was performed using the least-squares spectral analysis method of Lomb and Scargle that is appropriate for unevenly sampled data sets *(16,17,18)*. We chose to use the implementation in R by Glynn et al., adopting the p-values as introduced by Scargle for positive signal detection *(19)*. The resulting spectrum for the entire data set is depicted in Fig. 1 (bottom), where prominent peaks appear in both high and low frequencies. In the low frequency range, the dominant peak corresponds to the one year period that is clearly visible in the time series data. In the high frequency range, the scene is dominated by a triplet corresponding to the diurnal, semi-diurnal, and ter-diurnal peaks. The triplet was previously observed in other measurements where Fourier analyses were performed on uniformly sampled time series muon flux data *(20-25)*. The diurnal signal has the largest amplitude, but is only ~10% of that of the one year signal, as estimated from their power densities. Thus, the diurnal signal amplitude is approximately 0.5% of the average muon counts. As we integrate ~$10^5$ counts per hour, we are statistically bound to a precision of 0.3%, and hence statistics is the limiting factor in observing the diurnal signal unambiguously in the time series. To observe the signal's functional form, we averaged the signal amplitude for consecutive days, subtracting the mean number of counts for each day. This was done for each astronomical season in the 2010 data set, which, with ~90 days per season, results in a statistical accuracy of ~0.03%. The resulting experimental points are depicted in Fig. 2, bearing similarity to those obtained previously by other studies *(26-28)*. The red lines shown are the result of a fit performed with three sinusoidal functions with fixed 24, 12, and 8 *h* periods and amplitudes and phases that are searched for. The fitted parameters are listed in Table 1. The fitted curves describe the data well, with the striking feature that both phases and amplitudes have a strong seasonal dependence. The diurnal and semi-diurnal signals are stronger in Summer, whereas the ter-diurnal amplitude is larger in Winter.

A closer inspection of the individual peaks in the triplet reveals the presence of other peaks shown in Fig. 3. A "picket fence" like structure observed in other tidal spectral analyses is evident for all cases, with the fundamental frequency peaks in the center *(11,29)*. The ter-diurnal frequency spectrum was obtained by performing the least-squares spectral analysis on exclusively the winter months, while the diurnal and semi-diurnal frequency spectra were obtained by analyzing the remaining seasons. This procedure maximizes the signal to noise ratio for the respective frequency ranges. To extract the peak position, amplitude, and width, a fit assuming a normal distribution was applied to all peaks. The spectra are characteristic of a sinusoidal phase modulated signal with a carrier frequency centered at $\omega = n\, d^{-1}$ (*n = 1,2,3*) and a modulation frequency $\Omega = 0.0027\, d^{-1}$ (*1 yr*) found in the regular spacing between the peaks. Below each spectrum, we depict the peak position of previously observed tidal frequencies. Peaks that have corresponding p-values of less than $1\times10^{-4}$ are labeled by their Darwin indices *(30)*. Peaks that have local statistical significance, i.e. only within the pane presented, are labeled by their indices in parenthesis. The extracted peak position values are listed in Table 2, alongside those from the annual and semi-annual peaks. Previously published tidal frequencies are also tabulated *(11,30)*. On average, the agreement between this analysis and published values is within ± 0.0001 $d^{-1}$, well within the statistical uncertainties. No other peaks of significance were observed up to frequencies of 0.25 $d^{-1}$ (48 *h*). In particular, no gravitational tidal effects, lunar or solar, have been observed. The expected standard deviation for the number of samples acquired is ~$1.8\times10^{-4}\, d^{-1}$, which is consistent with our analysis. These observations, together with the seasonal dependence of the signal, favor the interpretation that the driving mechanism behind tides that modulate the cosmic ray flux is thermal in origin.

To gain further insight into the connection between atmospheric density and muon flux, we performed a spectral analysis of the atmospheric density at different altitudes. Twenty five kilometers east of the detector location, the Upton National Oceanographic and Atmospheric Administration (NOAA) station site launches sounding balloons at least twice a day that record measurements including temperature, pressure, and relative humidity. This data is publicly available through the Integrated Global Radiosonde Archive for measurements taken at altitudes up to 30 *km* (*31*). To perform a frequency analysis of atmospheric density as a function of altitude, twenty years of observational data were used to calculate average densities in 1 *km* slices. The data are an unevenly distributed time series due to variation in both launch time and launch frequency. If there were two regular samplings a day spaced by 12 *h*, the Nyquist frequency limit for reconstruction would be *1 $d^{-1}$*. However, the least squares spectral analysis of Lomb and Scargle applied to an unevenly sampled data set allows for this limit to be further extended, as aliasing is avoided by the fitting procedure*(32-34)*. Fig. 4 shows a set of frequency spectra as a function of altitude with those that have statistical significance, $p<1.0 \times 10^{-4}$, highlighted in red. Unique frequencies are seen both near sea level and at altitudes higher than 25 *km*, where strong atmospheric tides are experimentally observed. In a simple model where density variations elongate or contract the muon flight path, oscillations in density will induce muon flux modulations. Therefore, we hypothesize that these observed tidal frequencies reflect the modulations introduced by the atmospheric density variations at altitudes where mesons are produced. These are likely to be caused by the solar heating of the troposphere and the stratosphere. The study also shows that the density in the troposphere has a yearly periodic oscillation with a maximum amplitude in Summer and a minimum in Winter. The oscillation in the stratosphere has the same period as that in the troposphere, but is 180° out of phase, i.e. a larger amplitude in Winter. These amplitude variations are ±9% around the average density value of $4.4 \times 10^{-5}$ $g/cm^3$ at 25 km and ±6% around the average density value of $1.0 \times 10^{-3}$ $g/cm^3$ at 2 km. The effects of the latter variations are compensated for in the muon flux data by the barometric pressure correction.

The spectral features of Fig 1(bottom) were analyzed by performing simulation studies. Central to the studies is the role of missing data in the spectral reconstruction. A synthetic time series was constructed using both tidal signals and noise sources for the exact duration of the data collection period. The diurnal, semi-diurnal, and ter-diurnal signals were calculated using the procedure from reference 11. The diurnal signal was further modulated for seasonal variations; a yearly spaced sequence of inverted parabolic curves was added to reproduce the yearly flux modulation. Forbush deficit events of random amplitudes were randomly distributed throughout the time series using the event of February 2014 as a template. The two types of noise considered were the *$(1/f)^{-\alpha}$* spectral noise, commonly found in many physical systems, and the counting statistics, assumed to be given by the Poisson statistics, $\sigma = n^{1/2}$ *(35-38)*. For the latter, we took the average number of counts per hour, giving $\sigma = 5.6$ counts. Prior to frequency reconstruction, samples at the exact positions as the missing real data were removed from the time series. Results of the noise study are depicted in Fig. 5, where the experimentally obtained spectral noise is compared to three different simulation curves. For clarity, we present averaged values within a window of $\Delta f = 0.050$ *$d^{-1}$* and for $f > 5 \times 10^{-2}$ *$d^{-1}$*, which is free from any experimentally observed signals. It is found that $\alpha = 1.0$ satisfactorily reproduces the spectral noise for $f < 0.5$ *$d^{-1}$*. Above this frequency, the inclusion of the counting statistics seems to account for the deviation from the *1/f* spectral noise as seen in Fig 5, curve (b). Curve (c) shows the results of a simulation performed where a detector twenty times larger is assumed with all samples available. In this case, the departure from the *1/f* line is smaller. Curve (d) shows the results when only the *1/f* noise is considered. To show in detail the role of noise in the frequency reconstruction, Fig. 6 depicts the spectral lines for the diurnal frequency region for the cases discussed above. The two main effects of the small detector size and missing samples are the increase in the baseline spectral noise and the introduction of random small amplitude "spikes" near the peaks of interest. In these cases, the Scargle probabilities adequately flag those peaks that are artificially produced, with the caveat that weak signal peaks are also excluded.

We present evidence that tidal frequencies are observed in the spectral analysis of muon flux time series measurements. This result shows the intimate connection between the atmospheric density profile and particle production at the time a cosmic ray interacts with Earth's atmosphere. We suggest that the observed muon flux modulation results from atmospheric density variations due to thermal heating in the stratosphere and troposphere. The results of our simulation studies suggest that large area detectors would greatly enhance the sensitivity for the observation of frequencies larger than 0.5 $d^{-1}$. In the literature, we found at least two citations that seem to corroborate this conclusion. The GRAND experiment, composed of an array of muon detectors, can clearly observe periodicities of 6 $d^{-1}$ *(21)*. A second observation of waveforms, similar to those in Fig. 2 was made by the KEK experiment E391a, where the statistical uncertainty for each hour is already at the 0.03% level *(24)*. Tidal effects in the data seem to be already observed by the Tibet III experiment and at the HAWC observatory *(23,39)*. As cosmic rays have a known isotopic composition and energy spectra, shower simulation studies with the inclusion of atmospheric tides would be helpful in further understanding the transport of elementary particles through the atmosphere at different times of the year. For this it is highly desirable to describe the atmosphere profile using profile measurements such as those from GPS radio occultation techniques *(40)*. Cosmic ray shower reconstructions will benefit from these studies, as they are more sensitive to atmospheric density at points of particle interaction in the high atmosphere. These studies will also be helpful in understanding if geomagnetic tides have an effect on muon flux modulation. Modeling the cosmic ray transport in a realistic atmosphere will tell us how well elementary particles can be used to probe the atmosphere.

**Acknowledgments:** This work was supported in part by a NSF grant OCI-0636194 and by the United States Department of Energy Contract No. DE-SC0012704. We acknowledge the support from the Smithtown High School East in hosting and maintaining the scintillator telescope for over eight years of operation.

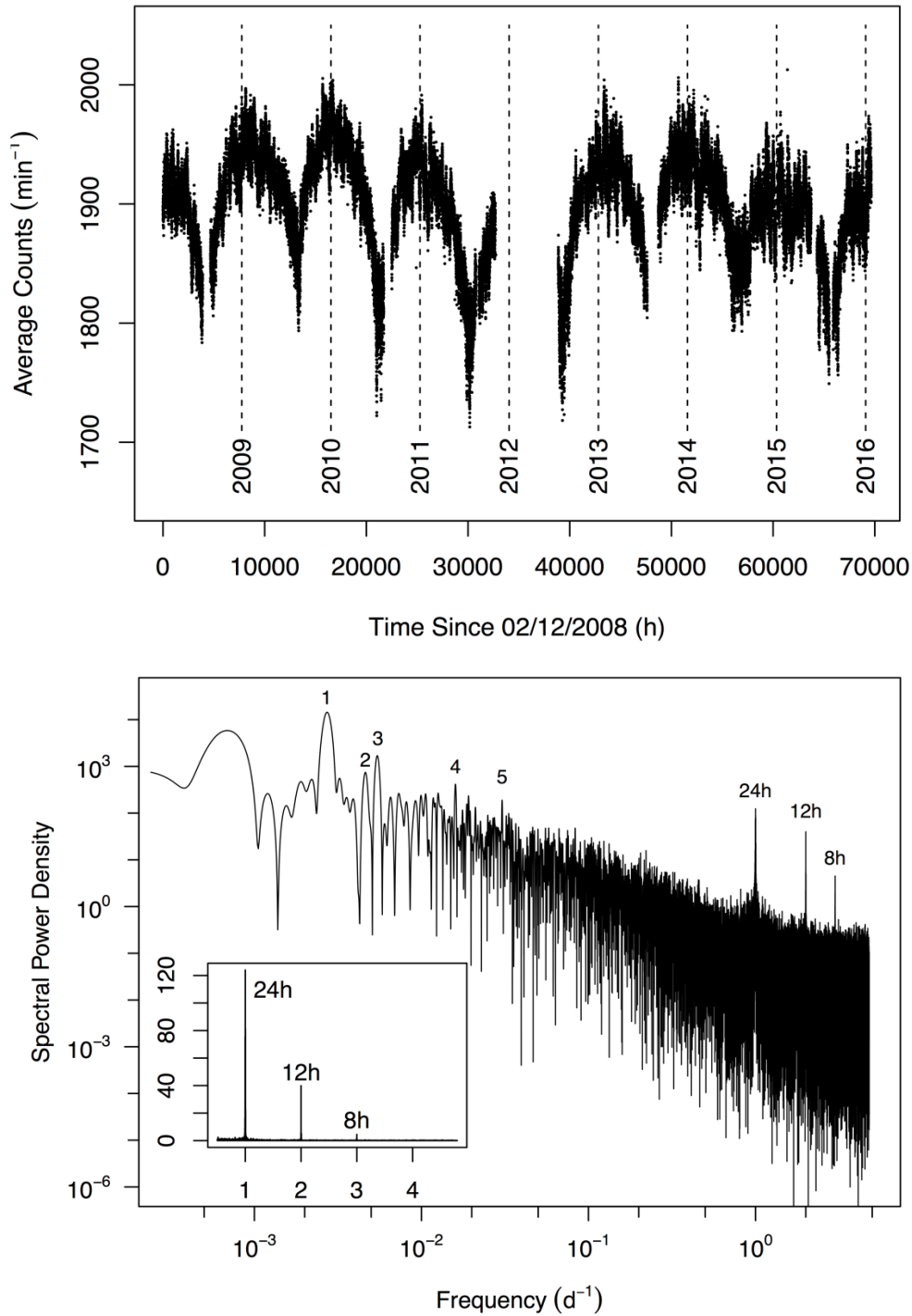

**Figure 1.** (**Top**) Hourly averaged counts displayed against the hour since the measurements have started in February of 2008. Counts were corrected for barometric pressure. (**Bottom**) Frequency spectrum of the time series obtained from the least square spectral analysis of Lomb and Scargle. The prominent peaks in the high frequency corresponding to 24 h, 12 h and 8 h periods are shown in the inset in linear scale. In the low frequency range we have identified five peaks. The peak corresponding to 1 year period , peak 1, is the strongest component. The frequencies for other peaks indicated by labels 2-5 are listed in Table 2.

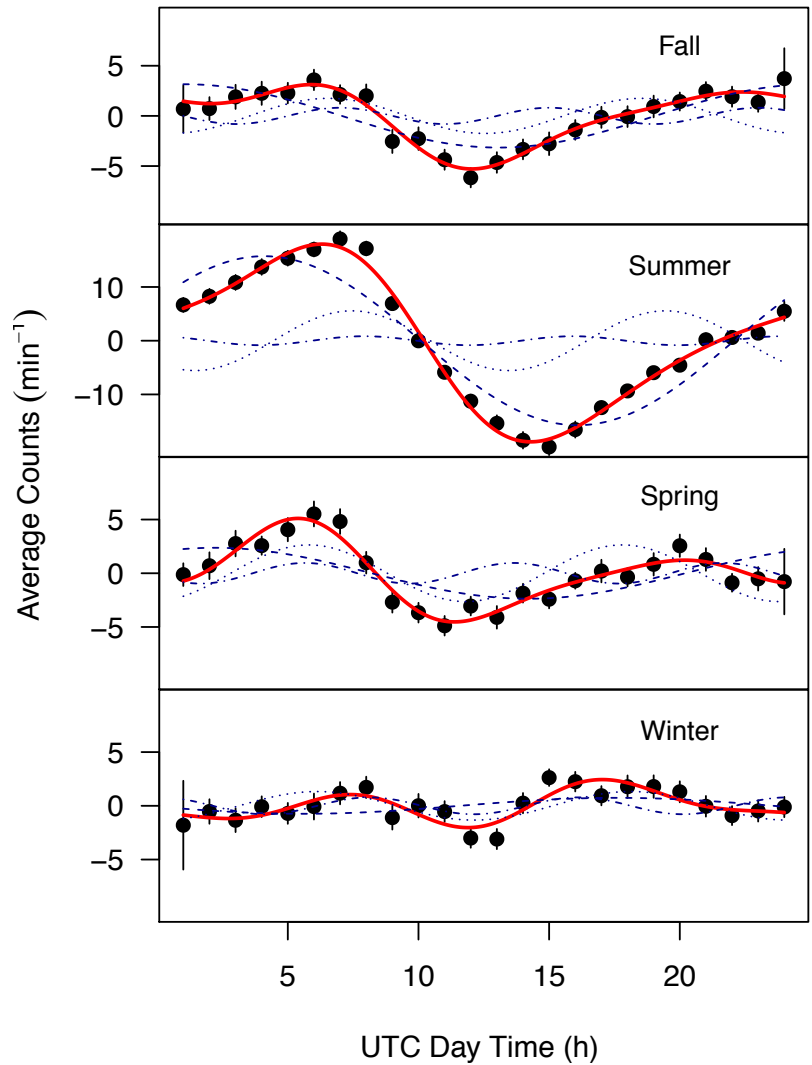

**Figure 2.** Signals corresponding to the tidal modulation for fall, summer, spring, and winter seasons, separately, in 2010. For each season, the counts for each hour were aligned and averaged after the mean value for each day was subtracted. The red curve is the result of a fit of three sinusoidal functions whose periods were fixed at 24, 12, and 8 *h*, and whose amplitudes and phases were adjusted for.

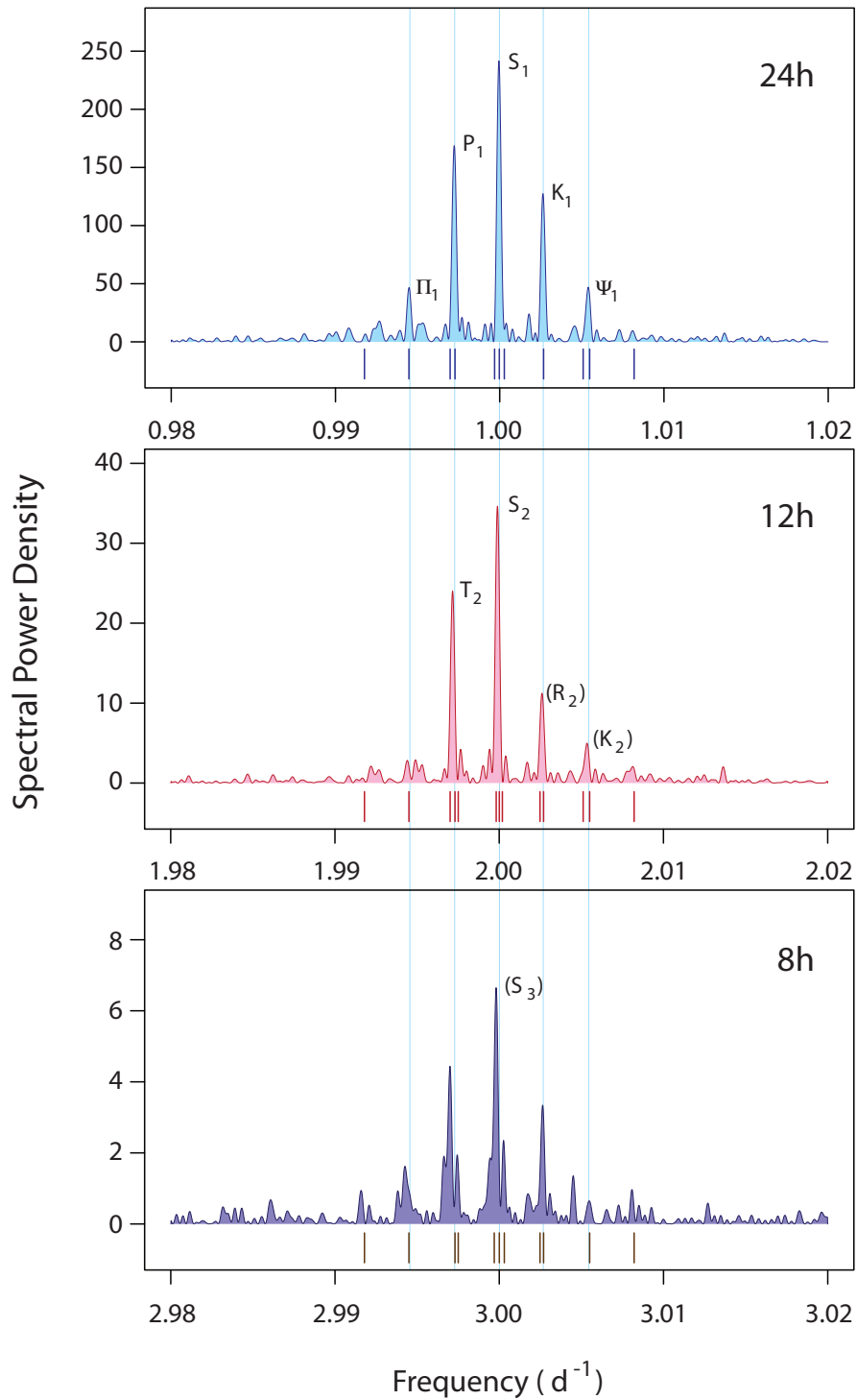

**Figure 3.** Frequency spectrum for the diurnal, semi-diurnal and ter-diurnal peaks. The known tidal frequencies are indicated in each window below the corresponding spectrum. The thin blue lines are the expected locations for the main tidal frequencies. Darwin indices for the peaks with p-values $p < 1 \times 10^{-4}$ are indicated. Peaks that have statistical significance only within the pane shown are indicated by their indices in parenthesis.

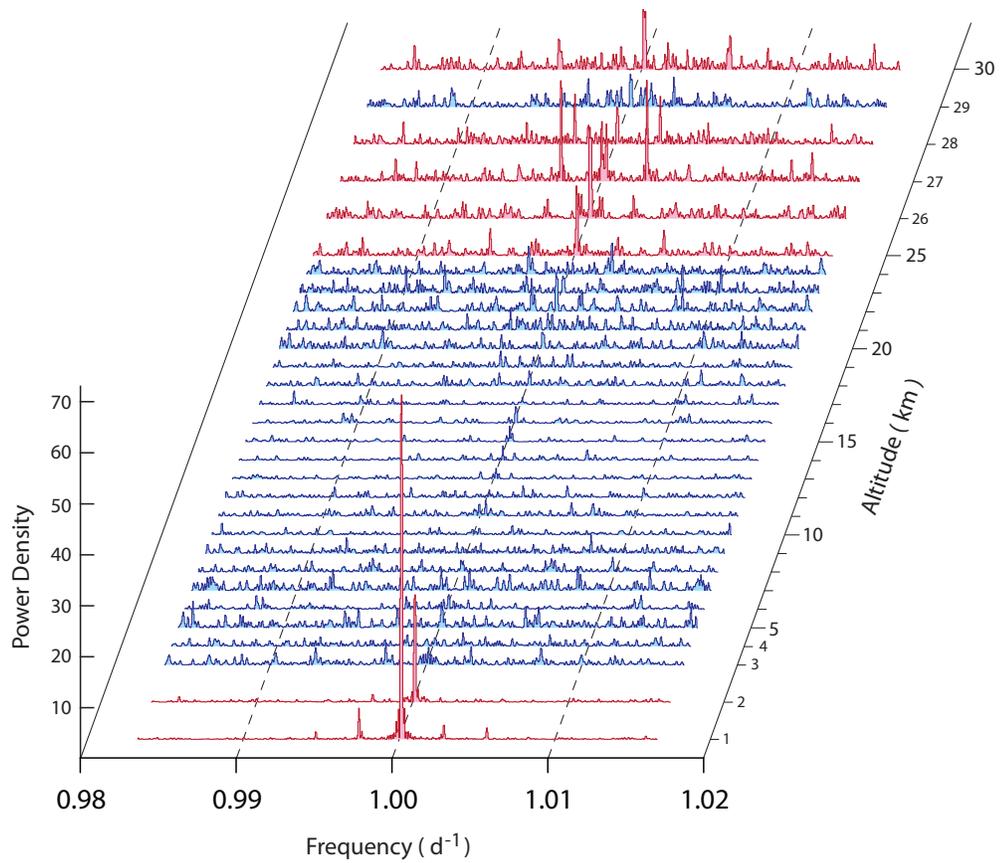

**Figure 4.** Frequency spectra for atmospheric pressure as a function of altitude. Spectra in red indicate the presence of peaks with statistical significance. Atmospheric densities were derived from balloon soundings at the Upton NOAA site taken over the past 20 years.

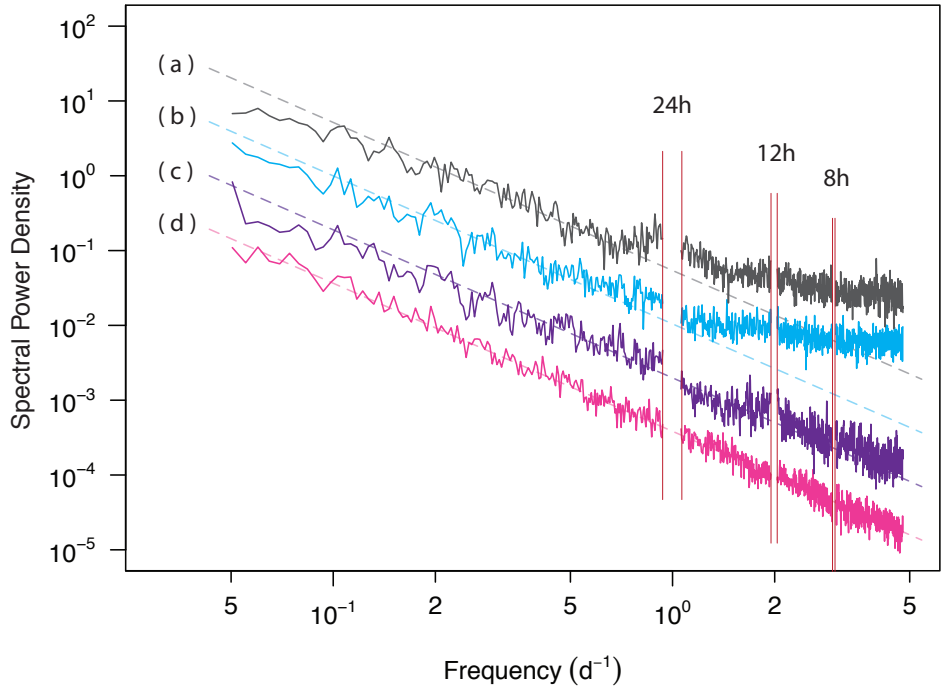

**Figure 5**. Averaged spectral noise compared to those obtained from simulation studies. The spectral noise was averaged in bins of 0.050 $d^{-1}$ in the range between $f = 5 \times 10^{-2}\ d^{-1}$ and $f = 5\ d^{-1}$. Each curve is offset by a factor of 8 from each other. Dashed lines represents the *1/f* noise for reference. The curve (a) is the experimental spectral noise and (b) obtained from simulation studies. The spectral noise in both cases is best described by the *1/f* noise with a clear upward turn observed for frequencies above ~0.5 $d^{-1}$. The deviation is found to be due to counting statistics, $\sigma = N^{1/2}$. Curve (c) is a simulation where this uncertainty is not included but missing samples are. Curve (d) is when only *1/f* noise is included in the simulation with all samples and no counting statistics considered. Regions around the signal of interest are excluded.

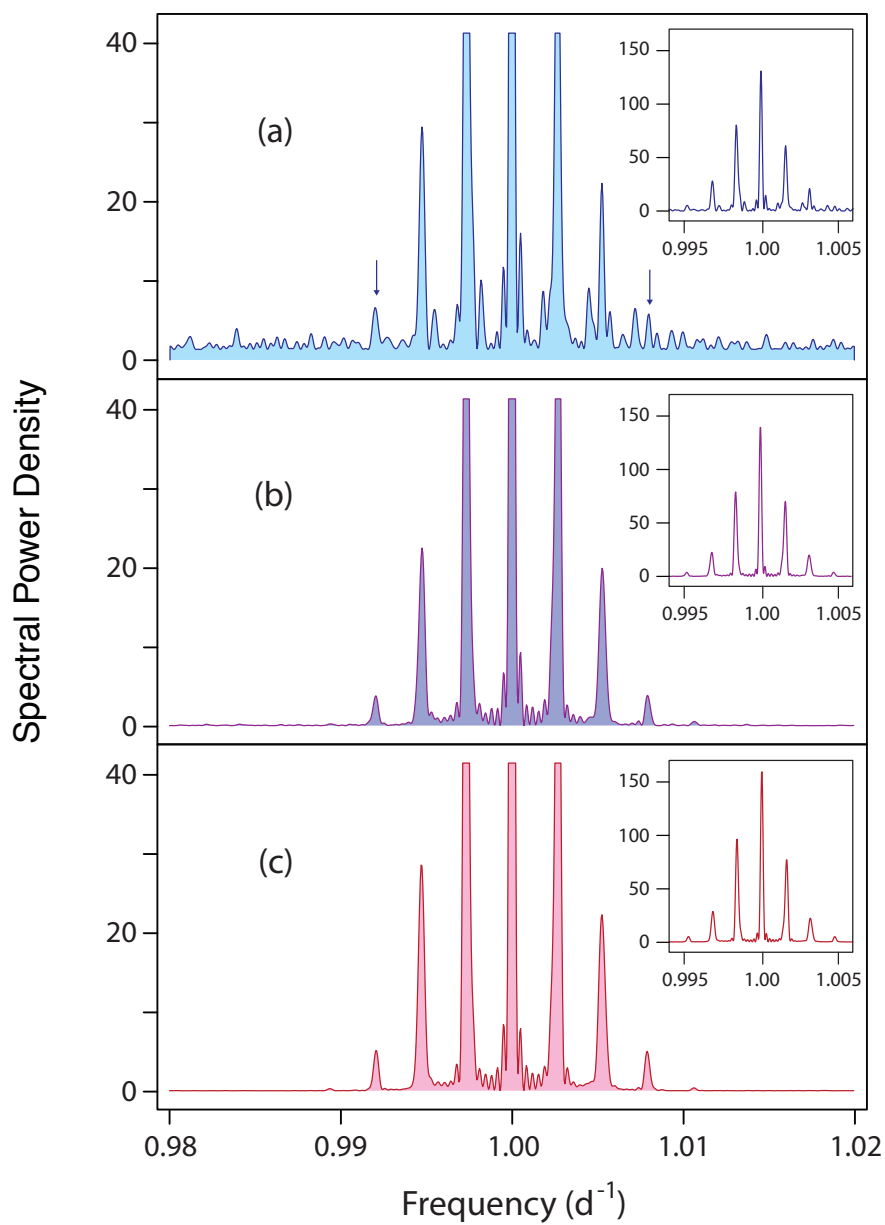

**Figure 6.** Diurnal frequency peaks obtained from the simulation studies for three different assumptions. (a) The 1/f noise, statistical uncertainties and missing samples are considered simulating our data collection conditions. (b) the 1/f noise, a large detector and all samples are considered, and (c) only the 1/f noise is included in the simulation. The small detector and missing samples increases the noise floor and introduce random peaks near the peaks of interest. The Scargle probabilities correctly assigns the false alarm probabilities for these peaks with the penalty that those with small spectral power, such as those indicated by arrows, are also tagged as noise.

| Season | $a_{24}$ | $\phi_{24}$ | $a_{12}$ | $\phi_{12}$ | $a_8$ | $\phi_8$ |
|---|---|---|---|---|---|---|
| Fall | 3.2±0.2 | -1.29±0.07 | 0.55±0.08 | 1.9±0.1 | 0.26±0.08 | -2.4±0.3 |
| Summer | 15.7±0.4 | 1.07±0.02 | 0.35±0.03 | 3.9±0.1 | 0.05±0.02 | -0.1±0.5 |
| Spring | 2.4±0.3 | -1.0±0.1 | 1.1±0.2 | 1.5±0.1 | 0.4±0.1 | 2.9±0.3 |
| Winter | 0.7±0.2 | 3.0±0.4 | 1.8±0.8 | 1.6±0.2 | 1.1±0.5 | -1.5±0.4 |

**Table 1**. Amplitudes and phases obtained from fits to the data points in Fig. 2. The fit function is the sum of three sinusoidal functions, $y = \Sigma \, a_n \sin(2\pi t/n + \phi_n)$ where $n = 24, 12$ and $8$. In the fit procedure the periods are fixed and the amplitude and phase are searched for.

| Peak | Frequency (d$^{-1}$) | Peak Power Density | (x10$^{-4}$) | p-value | Ref. 11 | Ref. 30 | Darwin Index |
|---|---|---|---|---|---|---|---|
| 1 | 0.0027 | 14469 | 1.8 | < 1x10$^{-11}$ | 0.0027 | — | $S_a$ |
| 2 | 0.0046 | 741 | 1.6 | <1x10$^{-11}$ | — | — | — |
| 3 | 0.0055 | 1706 | 1.8 | <1x10$^{-11}$ | 0.0055 | — | $S_{sa}$ |
| 4 | 0.0160 | 423.1 | 1.8 | <1x10$^{-11}$ | — | — | — |
| 5 | 0.0304 | 184.7 | 2.1 | <1x10$^{-11}$ | — | — | — |
| 6 | 0.9945 | 47.4 | 1.8 | <1x10$^{-11}$ | 0.9945 | 0.9945 | $\Pi_1$ |
| 7 | 0.9972 | 171.2 | 1.8 | <1x10$^{-11}$ | 0.9970 | 0.9973 | $P_1$ |
| 8 | 1.0000 | 245.3 | 1.7 | <1x10$^{-11}$ | 1.0000 | 1.0000 | $S_1$ |
| 9 | 1.0026 | 129.8 | 1.8 | <1x10$^{-11}$ | 1.0027 | 1.0027 | $K_1$ |
| 10 | 1.0054 | 47.6 | 1.9 | <1x10$^{-11}$ | 1.0055 | 1.0055 | $\Psi_1$ |
| 11 | 1.9972 | 24.4 | 1.7 | 3x10$^{-5}$ | 1.9973 | 1.9973 | $K_2$ |
| 12 | 2.0000 | 35.2 | 1.7 | 7x10$^{-10}$ | 2.0000 | 2.0000 | $S_2$ |
| 13 | 2.0026 | 11.4 | 1.8 | (0.0002) | 2.0027 | 2.0027 | $(R_2)$ |
| 14 | 2.0053 | 5.1 | 1.9 | (0.01) | 2.0055 | 2.0055 | $(K_2)$ |
| 15 | 2.9970 | 4.3 | 2.0 | (0.01) | 2.9973 | — | — |
| 16 | 2.9998 | 6.4 | 2.0 | (0.01) | 3.0000 | 3.0000 | $(S_3)$ |
| 17 | 3.0026 | 3.3 | 1.8 | (0.05) | 3.0027 | — | — |

**Table 2**. List of frequencies extracted from the spectral analysis of the muon flux time series measurement. For each identified peak, the frequency, peak spectral power, standard deviation, and p-value from the Lomb-Scargle analysis are listed. The first five rows are the peaks located in the low frequency region that are dominated by the 1 year and 1/2 year frequencies. All other peaks are located in the high frequency range. Peaks 6 through 12 have small p-values, and they are clearly above noise. Others have local significance and have estimated p-values for the pane in Fig 3. For comparison, the tabulated values from Ref 11 and 30 are listed together with their Darwin indices.